\documentclass{aa}

\usepackage{graphicx}
\usepackage{natbib}
\usepackage{txfonts}
\usepackage{color}
\usepackage[T1]{fontenc}
\usepackage{ulem}   

\def\lum{erg s$^{-1}$}

\def\aap{A\&A}

\def\apjs{ApJS}

\def\apjl{ApJLett}

\def\1v03{V~0332+53}
\def\4u{4U~0115+63}

\newcommand {\be}{\begin {equation}}
\newcommand {\ee}{\end {equation}}

\begin{document}

   \title{Propeller effect in two brightest transient X-ray pulsars: 4U~0115+63 and V~0332+53}

   \author{S.~S.~Tsygankov \inst{1}
          \and  A. A. Lutovinov \inst{2,3}
          \and V. Doroshenko \inst{4}
          \and A. A. Mushtukov \inst{5,6}
          \and V. Suleimanov \inst{4}
          \and J. Poutanen \inst{1,7}
          }

   \institute{Tuorla Observatory, Department of Physics and Astronomy,  University of Turku,
              V\"ais\"al\"antie 20, FI-21500, Piikki\"o, Finland \\ 
              \email{sergey.tsygankov@utu.fi}
       \and
             Space Research Institute of the Russian Academy of Sciences, Profsoyuznaya Str. 84/32, Moscow 117997, Russia
       \and
             Moscow Institute of Physics and Technology, Moscow region, Dolgoprudnyi, Russia
       \and
            Institut f\"ur Astronomie und Astrophysik, Universit\"at T\"ubingen, Sand 1, D-72076 T\"ubingen, Germany
       \and
            Anton Pannekoek Institute, University of Amsterdam, Science Park 904, 1098 XH Amsterdam, The Netherlands
       \and
           Pulkovo Observatory of the Russian Academy of Sciences, Saint Petersburg 196140, Russia
       \and
          Nordita, KTH Royal Institute of Technology and Stockholm University, Roslagstullsbacken 23, SE-10691 Stockholm, Sweden
       \
          }
   \titlerunning{Propeller effect in \1v03\ and \4u}
   \authorrunning{S. Tsygankov et al. }
   \date{Received; accepted}


  \abstract
  {}
   {We present the results of the monitoring programmes performed with the {\it Swift}/XRT telescope and aimed specifically to detect an abrupt decrease of the observed flux associated with a transition to the propeller regime in two well-known X-ray pulsars \4u\ and \1v03. }
   {Both sources form binary systems with Be optical companions and undergo so-called giant outbursts every 3-4 years. The current observational campaigns were performed with the {\it Swift}/XRT telescope in the soft X-ray band (0.5--10 keV) during the declining phases of the outbursts exhibited by both sources in 2015.}
{The transitions to the propeller regime were detected at the threshold luminosities of $(1.4\pm0.4)\times10^{36}$~erg~s$^{-1}$ and $(2.0\pm0.4)\times10^{36}$~erg~s$^{-1}$ for \4u\ and \1v03, respectively. Spectra of the sources are shown to be significantly softer during the low state.  In both sources, the accretion at rates close to the aforementioned threshold values briefly resumes during the periastron passage following the transition into the propeller regime. The strength of the dipole component of the magnetic field required to inhibit the accretion agrees well with estimates based on the position of the cyclotron lines in their spectra, thus excluding presence of a strong multipole component of the magnetic field in the vicinity of the neutron star.}
   {}


   \maketitle

%

\section{Introduction}

Interaction of matter with the magnetosphere of a neutron star
determines temporal and spectral properties of X-ray emission from
X-ray pulsars (XRPs), accreting millisecond pulsars (AMPs) and
recently discovered accreting magnetars. One of the most
straightforward manifestations of such an interaction is a transition of
the accreting neutron star to the so-called ``propeller regime''
\citep{1975A&A....39..185I}, when the accreting matter is stopped by
the centrifugal barrier set up by the rotating magnetosphere. Such
a transition is expected to occur at the limiting luminosity defined by
the dipolar magnetic field strength and rotation rate of the
pulsar. Two types of propellers with different properties can be
  defined as ``weak'' and ``strong'' \citep[see
    e.g.][]{2006ApJ...646..304U}. The strong propeller is characterized by
  the strong matter outflows and even formation of the collimated jets
  along the rotational axis of the neutron star. In both cases the
transition is accompanied by a dramatic drop of the X-ray flux
observed from the source. Confident detection of such a transition
provides a completely independent estimate of the magnetic field of
the pulsar and in general allows for a better understanding of  the interaction of
the accretion flow and magnetosphere. However, in spite of clear
importance of this effect, so far it has not been systematically
studied from observational point of view. One of the reasons is that
high sensitivity monitoring X-ray observations required to detect such
a transition only became possible recently.  Indeed, many accreting
pulsars exhibit strong variability, which might be interpreted as a
transition into propeller regime. For instance, it was demonstrated by
\cite{2011A&A...529A..52D,2012A&A...548A..19D} that the accretion is
not centrifugally inhibited during the so-called ``off-states''
observed in Vela~X-1 and 4U~1907+09, and despite significant
luminosity decrease these sources continue to accrete. On the other
hand, transition to the propeller regime is expected to inhibit
accretion almost completely, and in this case much lower luminosities
can be expected in the low state.

In fact, only a few cases of transitions to the propeller regime in X-ray
pulsars with known magnetic field strength were mentioned in the literature.
\cite{2008ApJ...684L..99C} associated the strong variations in the flux from a
weakly magnetized ($B\sim3\times10^8$~G) accreting millisecond pulsar
SAX~J1808.4--3658 with the onset of the propeller regime at the bolometric
limiting luminosity $L_{\rm lim}=(5\pm2)\times10^{35}$~erg~s$^{-1}$. Another
example is the bursting X-ray pulsar GRO J1744--28 with intermediate magnetic
field ($B\approx 5.3\times10^{11}$~G), which exhibited a similar rapid flux drop
at the threshold luminosity $L _{\rm lim}=(3.0\pm1.5)\times10^{37}$ erg
s$^{-1}$ \citep{1997ApJ...482L.163C}. Evidence for the
propeller effect was also reported for several weakly magnetized neutron stars
in LMXBs \citep[see e.g.][]{2013ApJ...773..117A,
2014MNRAS.441.1984C}. For a recent review, see \cite{2015SSRv..191..293R}.

Some hints of the transitions to the propeller regime in \4u\ and \1v03\ were
reported by \cite{2001ApJ...561..924C} and \cite{1986ApJ...308..669S}. The
former source was observed exactly during the transition when its luminosity
changed by factor of $\sim250$ during only $\sim15$ hours, but the
limiting luminosity $L_{\rm lim}$ was not measured. However, in the case of \1v03\
the transition was seen clearly at limiting luminosity $L_{\rm
lim}=(2.6\pm0.9)\times10^{36}$~erg~s$^{-1}$. Unfortunately, the low sensitivity of
{\it EXOSAT} did not permit to detect the source in the quiescent state. The
most strongly magnetized neutron star ($B\sim1\times10^{14}$~G) that demonstrates the
propeller effect is the first pulsating ULX M82 X-2 \citep{2014Natur.514..202B}
in which the centrifugal barrier stops the accretion below the luminosity $L_{\rm
lim}=2\times10^{40}$ erg s$^{-1}$ \citep{2016MNRAS.457.1101T}. A strong magnetic
field scenario is also supported by the detailed analysis of the accretion
column presented by \cite{2015MNRAS.454.2539M}, although other estimates based
on the analysis of the accretion torque acting onto the neutron star yield
field values in the much wider range of $10^{9}$ to $10^{14}$~G \citep[see e.g.][]{2015MNRAS.448L..43K, 2015MNRAS.448L..40E}.

In this work we present the results of the monitoring programmes performed with
the {\it Swift}/XRT telescope and aimed specifically to detect the
transition to the propeller regime and measurement of the quiescent flux in two well-known X-ray
pulsars \4u\ and \1v03. These are characterized by the spin periods of 3.6~s and 4.35~s and
magnetic fields of $B\approx 1.3\times10^{12}$~G
\citep{1983ApJ...270..711W} and $3.0\times10^{12}$~G
\citep{1990ApJ...365L..59M}, respectively. Both sources form binary systems
with Be optical companions (Be/XRP) and undergo the so-called giant outbursts every
3--4 years. The current observational campaigns were performed during the
declining phases of their outbursts in 2015 \citep{2015ATel.8179....1N,
2015ATel.7685....1N, 2015ATel.7822....1D}.

\section{{\it Swift}/XRT observations}

{\it Swift} observatory \citep{2004ApJ...611.1005G} permits us to
monitor sources of X-ray emission on very different timescales both in soft
and hard X-ray ranges. In this work we use observations of two Be/XRPs \4u\ and
\1v03\ performed with the XRT telescope \citep{2005SSRv..120..165B} covering the
declining phases of giant outbursts occurred in 2015. In particular, the
analysed data on \4u\ were collected between MJD 57329 and MJD 57388 and for
\1v03\ between MJD 57293 and MJD 57393.

The XRT observed both pulsars in photon counting (PC) and windowed timing
(WT) modes depending on their brightness. Final products (spectrum in
each observation) were prepared using the online tools provided by the UK
Swift Science Data Centre
\citep[\url{http://www.swift.ac.uk/user_objects/};][]{2009MNRAS.397.1177E}.

The obtained spectra were grouped to have at least one count per bin and
were fitted in {\sc XSPEC} package using Cash statistics
\citep{1979ApJ...228..939C}. To avoid any problems caused by the
calibration uncertainties at low
energies,\footnote{\url{http://www.swift.ac.uk/analysis/xrt/digest_cal.php}}
we restricted our spectral analysis to the 0.7--10 keV band.

\begin{figure}
\centering
\includegraphics[width=0.98\columnwidth, bb=20 230 570 675]{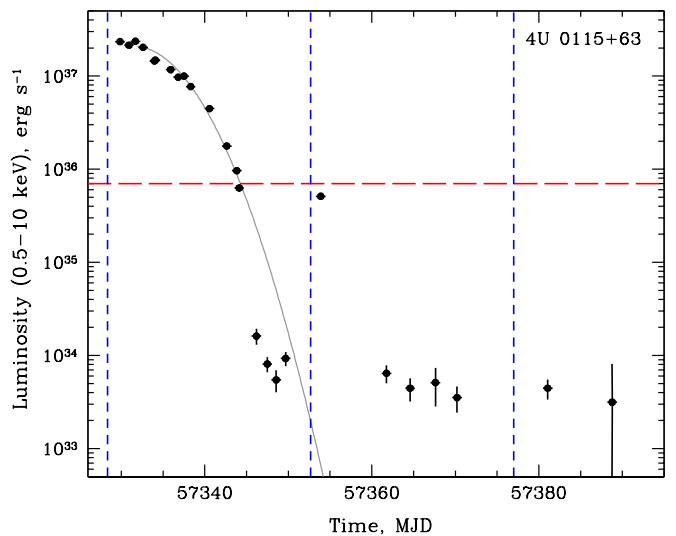}
\caption{Light curve of 4U 0115+63 obtained with the {\it Swift}/XRT
  telescope in the 0.5--10 keV energy range. Luminosity is calculated from the
  unabsorbed flux under the assumption of the distance $d=7$
  kpc. The solid grey line shows the best fit of the light curve before the transition to the propeller regime with a Gaussian function. The horizontal dashed line represents the
  limiting luminosity when the propeller regime sets in. Positions of vertical dashed
  lines correspond to the times of the periastron passage \citep{2010MNRAS.406.2663R}.
Applying the bolometric correction to the observed flux effectively increases the luminosity jump by factor of 2 (see Sec. \ref{sec:spec}).
}\label{fig:lc0115}
\end{figure}

\begin{figure}
\centering
\includegraphics[width=0.98\columnwidth, bb=20 230 570 675]{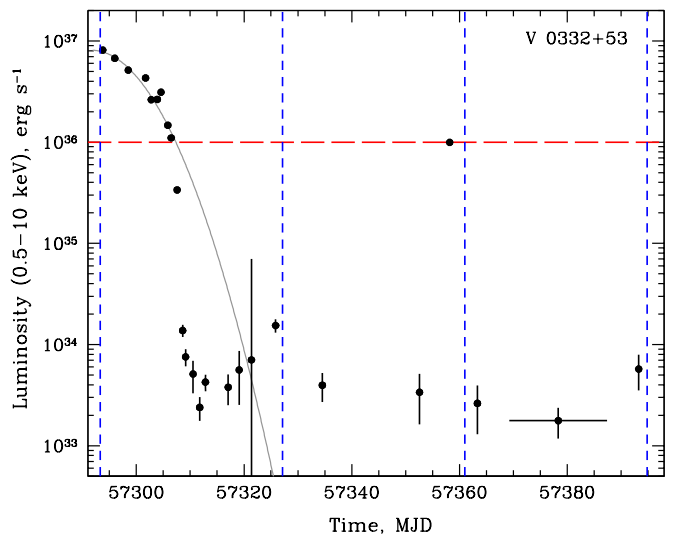}
\caption{Light curve of V 0332+53 obtained with the {\it Swift}/XRT
  telescope in the 0.5--10 keV energy range. Luminosity is calculated from the
  unabsorbed flux under assumption of the distance $d=7$ kpc. Point
  near the MJD 57380 was obtained from averaging of five observations with
  very low counting statistics. Solid grey line shows the best fit of the light curve before the transition to the propeller regime with a Gaussian function. Horizontal dashed line represents the
  limiting luminosity when the propeller regime sets in. Positions of vertical dashed
  lines correspond to the times of the periastron passage
  \citep{2016A&A...589A..72D}.
 Applying the bolometric correction to the observed flux effectively increases the luminosity jump by factor of 2 (see Sec. \ref{sec:spec}).
}\label{fig:lc0332}
\end{figure}

\section{Results}
\label{sec:res}

The light curves of \4u\ and \1v03\ as observed by {\it Swift}/XRT are shown
in Fig.~\ref{fig:lc0115} and Fig.~\ref{fig:lc0332},
respectively. The fluxes are given in the 0.5--10 keV energy range and
were corrected for the absorption. The main feature in both light curves is an
abrupt decrease of the observed flux below
$\sim7\times10^{35}$~erg~s$^{-1}$ for \4u\ and below
$\sim1\times10^{36}$~erg~s$^{-1}$ for \1v03\ (shown by red horizontal
dashed line). We interpret this behaviour as an onset of the
propeller regime (see Section \ref{sec:discus}).
We assume that the limiting luminosity associated with the transition to the
propeller regime corresponds to the dramatic change of the flux time derivative
(both figures have a logarithmic flux scale). The transition is
well illustrated by comparison of the light curve with the fit to its
bright part with a Gaussian function (shown with a solid grey line in
Figs.~\ref{fig:lc0115} and \ref{fig:lc0332}). The transition itself
can be fitted with an exponential function with e-folding times of
16.5 hr and 6.4 hr for \4u\ and \1v03, correspondingly. The bolometric correction increases the
  luminosity jump by factor of $\sim2$ (see Sec. \ref{sec:spec}) and
  hence decreases the e-folding times to 14.3 hr and 5.6 hr.
The blue vertical dashed lines represent times of the periastron passage,
according to the orbital ephemeris, obtained by \cite{2010MNRAS.406.2663R} for
\4u\ and \cite{2016A&A...589A..72D} for \1v03.

We estimated observed luminosities based on the fluxes
measured from the spectra, assuming distances to
both sources of 7\,kpc \citep{1999MNRAS.307..695N,2001A&A...369..108N}. We used two simple continuum models for
the
spectral fitting: the absorbed power-law model ({\sc phabs*powerlaw} in the {\sc XSPEC} package) and the
absorbed black-body emission model ({\sc phabs*bbody}). We found
that the power-law model approximates spectra of V\,0332+53 in the high
(accretion) state quite well with the photon index in the
range 0.4--0.6 and hydrogen column density $N_{\rm H} \simeq (1.0-1.3)\times10^{22}$
cm$^{-2}$.
The latter value slightly exceeds the interstellar
 absorption in the direction of the source derived from the Galactic
 neutral hydrogen maps, $N_{\rm H}\simeq0.7\times10^{22}$ cm$^{-2}$
 \citep{2005A&A...440..775K}.  
The counting statistics in an individual observation of V\,0332+53 in the low
state is not sufficient to determine the absorption independently. Therefore
to fit spectra in the low state we fixed it to the average value in the high
state of $1.2\times10^{22}$ cm$^{-2}$.

\begin{figure}
\centering
\includegraphics[width=0.98\columnwidth, bb=40 150 550 695]{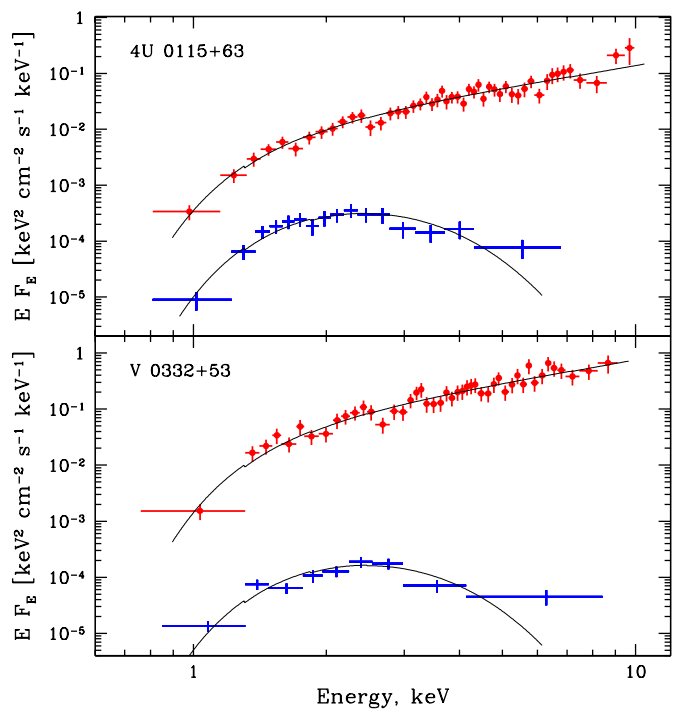}
\caption{ {\it Swift}/XRT spectra of \4u\ and \1v03\ in accretion (red dots)
  and propeller (blue crosses) regimes. Solid lines represent the best-fit models consisting of an absorbed power law in the high state and
  the absorbed black body in the low state.}\label{fig:specs}
\end{figure}

Also it is not possible to discriminate between different continuum
models in a single observation in the low state. To overcome these
difficulties we combined all observations with luminosities below
$1\times10^{34}$~erg~s$^{-1}$. The resulting spectrum has a much softer
shape in comparison to the bright state and can be well fitted ($C_{\rm stat} = 113.7$ 
for 132 d.o.f.) with the
absorbed black-body model with the temperature of $kT=0.51\pm0.03$\,keV and radius of the emitting area $R=0.6\pm0.1$ km. For comparison, the best-fit approximation with the power-law model 
gives the photon index $\Gamma=2.84\pm0.17$ and $C_{\rm stat} = 118.9$
for 132 d.o.f. In the subsequent analysis of spectra from all low-luminosity observations, we used a black-body continuum model 
 with the fixed absorption value and free black-body temperature. Analysis of the temperature evolution in the low
state is beyond the scope of this work and is presented elsewhere
\citep{2016arXiv160202275W}.  Here we would like to emphasize
the drastic change of the spectral shape due to the onset of the
propeller regime and cease of the accretion (Fig.\,\ref{fig:specs}).

The observed properties of 4U\,0115+63 are quite similar. At higher luminosities
its spectrum in the 0.5--10\,keV range is well described by a power-law model with the photon index in the range 0.4--0.8 and $N_{\rm
  H}\simeq(0.9-1.2)\times10^{22}$ cm$^{-2}$ (the interstellar value is
$N_{\rm H}\simeq0.86\times10^{22}$ cm$^{-2}$;
\citealt{2005A&A...440..775K}). To describe the average low-state
spectrum we used the absorbed black-body model with best-fit temperature of
$kT=0.52\pm0.03$ keV, radius of the emitting area $R=0.8\pm0.1$ km, the absorption value fixed at average value of $1.1\times10^{22}$
cm$^{-2}$, and $C_{\rm stat} = 118.9$ for 135 d.o.f. The best-fit approximation 
with the power-law model gives the photon index $\Gamma=2.70\pm0.17$ and $C_{\rm stat} = 138.6$
for 135 d.o.f. The difference in the source spectra in the high and
low states is apparently seen (Fig.\,\ref{fig:specs}).

\section{Discussion}
\label{sec:discus}
The sudden drop of the source luminosity in combination with the dramatic
spectral change strongly suggests that the accretion ceased in both sources,
which we interpret as a transition to the propeller regimes. Below we discuss this effect
in detail.

\subsection{Spectral changes}
\label{sec:spec}
As was described in Section \ref{sec:res}, the luminosity of both sources
in low state is a ${\rm few}\times10^{33}$~erg~s$^{-1}$ and their spectra are
compatible with the black-body model with temperature around 0.5
keV.

Using the Stefan-Boltzmann law we can estimate the area needed to
explain the observed flux, \be S=L/(\sigma_{\rm SB}T^4)\simeq
1.8\times 10^{10} L_{34}/T^{4}_{\rm keV}\,\,\,{\rm cm^2}.  \ee
Considering the observed temperature $T\simeq0.5$~keV, this gives an
area that is comparable with the area of polar cups at the surface of accreting
neutron star.  Taking into the account that observations took place
right after a major outburst, it is natural to associate the observed
soft emission with radiation from the cooling polar cups
\citep{2016arXiv160202275W}.  Indeed, \cite{2007ApJ...658..514R}
estimate the luminosity of a cooling neutron star at $\sim0.5$\,\% of
the average outburst flux, which is fully consistent with observations
for both \4u\ and \1v03.

On the other hand, residual accretion also cannot be excluded. Indeed, the
typical hard spectrum observed from X-ray pulsars might in principle become significantly
softer in quiescence. The hard power-law tail is formed because of the
Comptonization of initially thermal X-ray spectrum by electrons in the
optically thin atmosphere \citep{1969SvA....13..175Z} and further in the
accretion channel \citep{1982SvAL....8..330L,2007ApJ...654..435B}. The spectral
changes in the atmosphere are comparatively small in the case of low mass
accretion rates
\citep{1969SvA....13..175Z}. The Thomson optical thickness across the accretion channel
$\tau \approx 2\,L_{37}^{6/7}B_{12}^{2/7}$ \citep{2015MNRAS.447.1847M} is well
below unity for accretion luminosity $\sim 10^{34}-10^{35}\,{\rm erg\,s^{-1}}$.
Therefore, at low accretion rates  Comptonization in the accretion channel
is not expected to significantly affect the initial spectrum as well,
so it is likely to remain relatively soft.
The variations of spectral hardness and its correlation with the mass accretion
rate for XRPs were discussed recently by \cite{2015MNRAS.452.1601P} 
in context of a reflection model (see \citealt{2013ApJ...777..115P}).

Regardless of the origin of the emission in the low (propeller) state,
we can use our spectral fits to estimate the bolometric corrections
for both intensity states, which is important to constrain the
bolometric transitional luminosity and thus accretion rate.  For this purpose we used the results of our
spectral analysis of the emission of the sources in a wide energy range,
 obtained during previous outbursts with the {\it RXTE} and {\it
  INTEGRAL} observatories
\citep{2006MNRAS.371...19T,2007AstL...33..368T}. It was found that the
source flux in the 0.5--10 keV energy was $\sim50$\% from the
bolometric (i.e. flux in the 0.5--100 keV energy range). Based on
these results, in the subsequent calculations we used the coefficient
2 for the conversion of fluxes for both sources in the bright (accretion)
state.  In the propeller state, {\it Swift}/XRT energy coverage is
sufficient to detect most of the emergent flux and no additional
bolometric correction is needed. Indeed, this conclusion is 
valid both for the black body and power-law models due to the
low temperatures ($kT\simeq0.5$ keV) or steep photon indexes ($\Gamma\simeq2.7-2.8$).

The bolometric correction may introduce
additional uncertainty because of our ignorance of the broadband source
spectrum at luminosities around the transition. We estimated this uncertainty at
the level of 10--20\% and added it to the final uncertainty for the
limiting luminosity. Taking all uncertainties into account, we estimate the
bolometric limiting luminosities for the onset of the propeller regime at
$(1.4\pm0.4)\times10^{36}$~erg~s$^{-1}$ and
$(2.0\pm0.4)\times10^{36}$~erg~s$^{-1}$ for \4u\ and \1v03, respectively.

\begin{figure}
\centering
\includegraphics[width=0.98\columnwidth,bb=60 275 565 680]{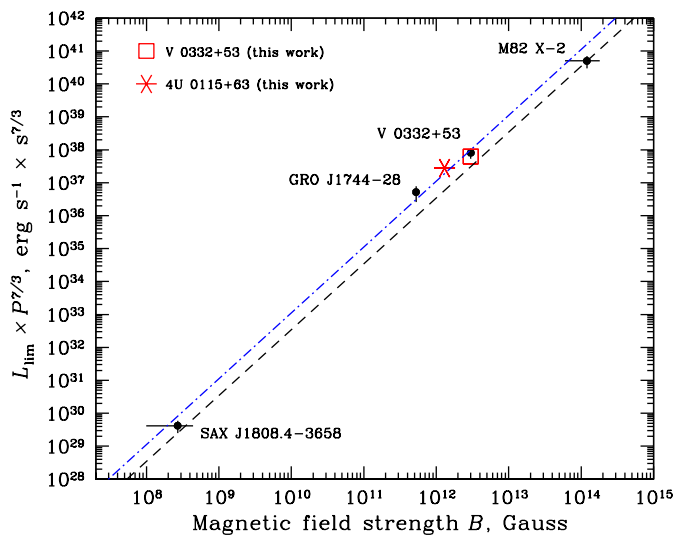}
\caption{Observed correlation between a combination of the
  propeller limiting luminosity and the period, $L_{\rm lim}P^{7/3}$,
  and independently determined magnetic field strength $B$ for five
  pulsars and one accreting magnetar (M82 X-2) is shown with circles
  with error bars \citep{2016MNRAS.457.1101T}.  The dashed and dash-dotted lines represent the
  theoretical dependence from equation~(\ref{eq1}), assuming NS
  standard parameters ($M=1.4 M_{\odot}$, $R=10$~km) and $k=0.5$ and
  $k=0.7$, respectively.  The red star and open box indicate the positions
  of \4u\ and \1v03\ obtained in this work, respectively.
}
\label{fig:collect}
\end{figure}

\subsection{``Propeller'' effect}

The ``propeller effect'' \citep{1975A&A....39..185I} is caused by a
substantial centrifugal barrier produced by the rotating magnetosphere
of the neutron star. The matter stopped at the magnetospheric radius
can only penetrate the magnetosphere if the velocity of the field
lines is lower than the local Keplerian velocity. In other words, the
matter will be accreted only in the case when the magnetospheric radius
($R_{\rm m}$) is smaller than the corotation radius ($R_{\rm c}$). The
magnetospheric radius depends on the mass accretion rate, and,
therefore, one can estimate the limiting luminosity corresponding
to the onset of the propeller regime by equating the corotation and magnetospheric radii
\citep[see e.g.][]{2002ApJ...580..389C}
\be\label{eq1}
L_{\rm lim}(R) \simeq \frac{GM\dot{M}_{\rm lim}}{R}
\simeq 4 \times 10^{37} k^{7/2} B_{12}^2 P^{-7/3} M_{1.4}^{-2/3} R_6^5 \,\textrm{erg s$^{-1}$} ,
\ee
where $M_{1.4}$ is the neutron star mass in units of 1.4M$_\odot$,
$R_6$ is neutron star radius in units of $10^6$~cm, $P$ its rotational
period in seconds, $B_{12}$ is the magnetic field strength in units of
$10^{12}$~G, and $\dot{M}$ is the mass accretion rate onto the neutron
star.  Factor $k$ relates the magnetospheric radius for
disc accretion to the Alfv\'en radius calculated for spherical
accretion and is usually assumed to be $k=0.5$ \citep{GL1978}.

Because the accretion efficiency in the propeller regime is much
lower, the decrease of the accretion rate below $\dot{M}_{\rm lim}$
leads to the drop in the luminosity to the value expected for the so-called ``magnetospheric accretion'' case \citep{1996ApJ...457L..31C}
\be\label{eq2}
L_{\rm lim}(R_{\rm c}) = \frac{GM\dot{M}_{\rm lim}}{R_{\rm c}}
= L_{\rm lim}(R) \frac{R}{R_{\rm c}} .
\ee

Therefore, in the absence of other sources of emission in the low
(propeller) state we can expect a very large drop of the observed
luminosity
\be\label{eq3}
\frac{L_{\rm lim}(R)}{L_{\rm lim}(R_{c})} = \left(\frac{GMP^2}{4\pi^2R^3}\right)^{1/3}
\simeq 170 P^{2/3} M_{1.4}^{1/3} R_6^{-1} .
\ee

Substituting spin periods of \4u\ and \1v03\ to equation (\ref{eq3}), we get the
expected drop in luminosities of $\sim400$ and $\sim450$, respectively.
However, these estimates do not take into account a substantial shift of the spectral distribution in the case of magnetospheric accretion. Indeed, the temperature of the accretion disc at the magnetospheric radius is
\be
T_{\rm keV}(R_{\rm m})\simeq 0.03\,B_{12}^{-3/7}L_{37}^{13/28}.
\ee
For the mass accretion rate corresponding to the transition to the propeller
regime for both pulsars, (about $10^{16}\,{\rm g/s}$) the disc temperature at
$R_{\rm m}$ is $\sim 10\,{\rm eV}\approx 10^5\,{\rm K,}$ which corresponds to
emission in the UV band for the magnetospheric accretion
\citep{1994ApJ...423L..47S}.

Therefore, the drop in X-ray luminosity that is calculated using equation (\ref{eq3})
should only be considered as a lower limit. On the other hand, additional
emission might come directly from the neutron star either owing to the leakage of
matter through the centrifugal barrier or thermal emission of the cooling NS
surface (Sect. \ref{sec:spec}). Taking into the account that interaction of the
accretion disc with the magnetosphere is not fully understood, and the likely
additional contribution to the flux in the low state, we consider the observed
drops in the luminosity ($200\pm50$ and $250\pm50$ for \4u\ and \1v03,
respectively) to be in reasonable agreement with theory. 

Another expected consequence of the onset of the propeller regime is a
substantial decrease of the pulsed fraction owing to partial or even complete
blocking of the accretion along magnetic field lines. Unfortunately, we were
unable to verify this assumption because of the insufficient temporal resolution of
XRT telescope in the PC mode (2.5 s) and, more importantly, extremely low flux
from both sources in the propeller state resulting in a count rate about
$10^{-2}$~cnt~s$^{-1}$ or less.

Figure~\ref{fig:collect} shows the correlation between the
independently measured magnetic field strength and combination of the
propeller limiting luminosity and the period, $L_{\rm lim}P^{7/3}$,
for four sources exhibiting transitions to the propeller
regime. Measurements taken from the literature  \citep[for details see][]{2016MNRAS.457.1101T} are shown with circles
with error bars (from \citealt{2009MNRAS.400..492I}, for SAX
  J1808.4--3658; \citealt{2015MNRAS.449.4288D} and \citealt{2015MNRAS.452.2490D},
  for GRO J1744--28; \citealt{1983ApJ...270..711W}, for \4u;  \citealt{1990ApJ...365L..59M}, for \1v03; \citealt{2015MNRAS.454.2539M}, for
  M82~X-2). The dashed
line represents the theoretical dependence from equation~(\ref{eq1})
assuming the standard parameters $M=1.4 M_{\odot}$, $R=10$~km,
$k=0.5$. The star and open box indicate the positions of \4u\ and
\1v03\ obtained in this work, respectively.  Good agreement with the
theoretical prediction can be clearly seen. Taking into account the strong
dependence of the limiting luminosity $L_{\rm lim}$ on the parameter
$k,$ we can roughly restrict the range of its values between 0.5 and
0.7 (shown by dashed and dash-dotted lines, respectively).

\subsection{Episodes of resumed accretion}
Light curves of \4u\ and \1v03,\ in addition to the main propeller transition,
exhibit another interesting feature: in both cases a sudden increase of the
bolometric flux is observed (on MJD~57354 up to $1\times10^{36}$ \lum~and on MJD~57358 up
to $2\times10^{36}$ \lum\ , respectively). The corresponding luminosities are within the error
consistent with the limiting luminosities for the propeller regime in \4u\
($L_{\rm lim}=(1.4\pm0.4)\times10^{36}$~erg~s$^{-1}$) and \1v03\ ($L_{\rm
lim}=(2.0\pm0.4)\times10^{36}$~erg~s$^{-1}$). We interpret this brightening as
a temporary transition back to the accretor regime due to increased mass
accretion rate.

Indeed, in both cases the re-brightening occurred close to the periastron passage
(see Figs. \ref{fig:lc0115} and \ref{fig:lc0332});  such episodes of
intermittent accretion are expected for the case of steady low-level mass
accretion \citep[see
e.g.][]{2010MNRAS.406.1208D,2012MNRAS.420..416D,2013A&A...550A..99Z}.

Spectra of both sources significantly harden during the resumed accretion
episodes becoming consistent with the spectrum right before the first
transition to the propeller. We note that the regular transitions from the
accretor to propeller and back are observed in another X-ray pulsar --
the accreting magnetar M82~X-2 \citep{2016MNRAS.457.1101T}.

\subsection{Accretion disc instability}

Exponential decay of the outbursts is not a unique property of X-ray pulsar
giant II-type outburst. Many other classes of outbursting astrophysical
objects, such as dwarf novae and soft X-ray transients, show similar flux decay
in the end of the bursts. Current interpretation is that the accretion
disc instability arising from partial ionization of the hydrogen (so-called
thermal-viscosity instability) governs the observed outburst dynamics
\citep{1984A&A...132..143M,1984AcA....34..161S,1993PASJ...45..707M,
1992ApJ...397..664C,1997ApJ...491..312C}. Here we discuss whether disc 
instability can be responsible for the observed behaviour of \4u\ and \1v03. 

Ionization of disc plasma implies significant changes of the opacity and gas
equation of state, and, therefore, $\alpha-$viscosity (see e.g. the review by
\citealt{2001NewAR..45..449L}). This leads to a drastic
change in the accretion rate and defines a general picture of a
dwarf nova outbursts. 

There are two different states of the accretion disc: the cold state with
mainly atomic hydrogen and low viscosity, when the accreted matter
from the secondary accumulated in the disc, and the hot state with mainly
ionized hydrogen and high viscosity, when all the accumulated matter rapidly
accretes on the central object (white dwarf in the case of dwarf nova).
The transition between the states occurs at the local effective temperature of $\sim 6500$\,K.
The accretion process is stable if the temperature is below the critical value over the entire disc,
i.e. at the sufficiently low mass accretion rate \citep{1997ASPC..121..351L},
\be
\dot{M}\lesssim 3.5\times 10^{15} r_{\rm in,10}^{2.65}M_{1.4}^{-0.88}\,\,{\rm g\,s^{-1}}, 
\ee
or if the temperature is higher than the critical temperature over the entire disc. This corresponds to the situation when mass accretion rate is high enough,
\be
\dot{M}\gtrsim 6\times 10^{16} r_{\rm out,10}^3\,\,{\rm g\,s^{-1}},
\ee
where $r_{\rm in,10}=r_{\rm in}/10^{10}\,{\rm cm}$ is the inner disc radius and
$r_{\rm out,10}=r_{\rm out}/10^{10}\,{\rm cm}$ is the outer disc radius. Intermediate mass accretion rate leads to developing the instability. 

Initial transition from the cold to hot state occurs at the inner or 
outer part of the disc \citep{1984AcA....34..161S}. The transition is triggered
by a heating wave propagating through the disc. The transition wave velocity ${\rm v}_{\rm tr} \sim \alpha
{\rm c}_{\rm s} (H/r)^{0.5}$ \citep{1996ApJ...471..921V} is a few times higher
than the radial velocity of the matter in the disc ${\rm
v}_{\rm R} \sim \alpha {\rm c}_{\rm s} H/r$, where ${\rm c}_{\rm s} \approx
10^6\,(T/10^4 \,{\rm K})^{1/2}$\,cm\,s$^{-1}$ is the local sound speed, $\alpha
\sim 0.01 -0.1$ is a viscosity parameter, and $H$ is a local disc
half thickness. We assume $\alpha=0.1$ for all subsequent estimates. The reverse
transition to the cold state happens when the disc temperature drops back to
the critical temperature at some radius. At this stage the cooling front arises and
propagates inwards from larger to smaller radii. Assuming dimensionless viscosity $\alpha\propto (H/r)^n$ with
$n\simeq 1.5$ the velocity of the cooling front $v_{tr}\propto r$
\citep{1995ApJ...454..880C,1996ApJ...471..921V}. 
As a consequence the hot inner parts lose angular momentum and the mass transfer rate decays exponentially at any radii
(including the inner disc radius). The accretion rate thus decays exponentially on timescales that are sufficient for the cooling
front to reach the inner disc radius.
 
The disc viscosity diffusion time $\sim r_{\rm out}/{\rm v}_{\rm R}(r_{\rm
out})$ determines the outburst duration, whereas the timescales for rising and declining phases are determined by the transition wave velocity
$\sim r_{\rm out}/{\rm v}_{\rm tr}(r_{\rm out})$. Observed rising and declining
times can be shorter because the accretion luminosity is determined by
processes that are near the inner disc radius.

Dwarf novae are, as a rule, compact systems with orbital periods $P_{\rm orb}$
of roughly a few hours, and their discs are also compact with $r_{\rm out} \sim
10^{10}$\,cm. As a result, their outbursts are relatively short ($\sim
10-20$\,days) with the rising and  declining time lasting roughly a few days
\citep{2003cvs..book.....W}. Because of the relatively small
size of the accretion discs, the averaged mass transfer rate can be high enough to keep entire
disc in a hot state. Such systems in the permanent outburst are known as
nova-like variables \citep{2003cvs..book.....W}.
 
The binary system consisting of V~0332+53 and its optical companion BQ Cam is not compact: the orbital
period is $P_{\rm orb} \approx 34$ days and the distance between companions
varies from $6.3\times 10^{12}$ to $8.5\times 10^{12}$\,cm
\citep{1999MNRAS.307..695N} because of the neutron star orbit eccentricity. The
tidal radius of the neutron stars Roche lobe is $\sim 10^{12}$\,cm and the
effective temperature at this radius is less than 1000\,K even for the maximum
mass accretion rate $\dot M \sim 10^{18}$\,g\,s$^{-1}$ at the peak of the
outburst. Therefore, it is probable that the disc instability takes place
in the system.

The total mass of the steady-state disc in outburst can be estimated by
integrating the surface density $\Sigma(R)$
from the corotation radius, which is $\sim 4.6\times 10^8$\,cm, up to the
radius where the effective temperature reaches critical temperature of
6500\,K, which is $\sim 7\times 10^{10}$\,cm for the mass accretion rate $\dot
M = 10^{18}$\,g\,s$^{-1}$ and the neutron star mass 1.5 $M_\odot$,
\be
   M_{\rm d} \approx 2\pi \int_{R_{\rm c}}^{r_{\rm out}}\,\Sigma(r)\,r\,dr \approx 7\times 10^{24}\,\alpha_{0.1}^{-4/5}\,{\rm g}, 
   \ee 
where $\alpha_{0.1} \equiv \alpha/0.1$. This is the lower limit of the total
disc mass because a larger part of the disc can be involved in the outburst
due to irradiation. Here we use the
steady-state $\alpha-$disc model \citep{1973A&A....24..337S} without irradiation,
using the analytic solution for zone C, which is presented by \cite{2007ARep...51..549S}.

On the other hand, total fluence of the current
outburst was $E_{\rm fl} \approx 10^{45}$\,erg and, therefore, the total accreted
mass is $\Delta M = E_{\rm fl}/(0.1c^2)\approx 10^{25}$\,g. This value is close
to the estimated total mass of the disc suggesting that most of the disc was, in fact,
accreted during the outburst. The typical viscous time, $r_{\rm out}/v_{\rm R} \sim
10^{11}/10^4 \approx 10^7$\,s, is also comparable to an order of magnitude
with the observed outburst decay timescale. The outburst evolution is, therefore,
indeed quite similar to that observed in dwarf novae.

The situation is more complicated, however, for a magnetized neutron star.
If the magnetospheric radius during the quiescence remains smaller than
the corotation radius, the flux decay is caused by the disc transition into
a cold state, which is similar to the outburst declining in non-magnetized systems. This
case is demonstrated by dwarf nova GK~Per with a magnetized white dwarf as a
main star \citep{2016arXiv160400232S}. If this is not the
case, however, the accretion might be inhibited before the transition cooling wave
reaches the corotation radius.

One can estimate typical timescales for mass accretion rate decay in both
cases. The fast transition to the quiescence occurs at the luminosity of about
$10^{36}$\,erg\,s$^{-1}$ for V~0332+53. The corresponding outer disc radius,
where the effective temperature equals 6500\,K, is $\sim 1.5\times 10^{10}$\,cm
and $H/r \approx 0.04$. Therefore, the transition wave velocity is $\sim
10^4$\,cm\,s$^{-1}$ and the typical decay time from the disc instability
has to be $\sim 10^6$\,s, i.e. $\sim 10$\,days. On the other hand, the accretion quenching due to
the propeller effect has to be realized near the corotation radius with a
typical diffusion velocity, which is, in this case about $1.5\times10^4$\,cm\,s$^{-1}$.
The corresponding timescale is then $\sim
3\times10^4$\,s, or $\sim 8$ hours. As it was shown before, the typical fast
decay time is $\sim 7$ hours for V~0332+53. Therefore, we conclude that the
propeller effect is more probable cause for the accretion quenching in this case. 
The main conclusions would be the same for our
second object, 4U~0115+63.

The accretion disc irradiation is significant during the outburst. It can
potentially affect the outer disc properties and keep the disc
in a hot stable state longer (see e.g.
\citealt{1996ApJ...464L.139V,1998MNRAS.293L..42K}). The
irradiation also makes the typical timescales of disc instability
evolution even longer. In this sense the main conclusion about the nature of observed
luminosity decay remains the same.

\section{Conclusion}

In this work we presented the results of the monitoring programmes
dedicated to detecting the transitions to the propeller regime in two
bright transient X-ray pulsars, \4u\ and \1v03,\ during their recent giant
outbursts in 2015. In both sources such transitions were detected with
confidence for the first time. The threshold luminosities for the
propeller regime onset are $L_{\rm
  lim}=(1.4\pm0.4)\times10^{36}$~erg~s$^{-1}$ and $L_{\rm
  lim}=(2.0\pm0.4)\times10^{36}$~erg~s$^{-1}$ for \4u\ and \1v03,
respectively. The luminosity drop by a factor of $\sim200$
and $\sim250$ is somewhat lower than expected
($\sim400$ and $\sim450$), which suggests the existence of some additional
source of emission in the low (propeller) states in both sources. The
most probable source of this emission is the cooling of the neutron star
surface heated during the outbursts. The observed substantial
softening of the source spectrum in the low state in combination with
the estimated emitting area support this supposition.

Both sources exhibit brief episodes of resumed accretion as the
quiescent neutron star approaches the periastron. Interestingly, the
luminosity during these episodes coincides with the limiting
luminosity for the propeller transition. The spectrum also becomes hard again
suggesting the revival of the accretion onto the neutron star surface.
Similar transitions are observed regularly in the first pulsating
ULX~M82~X-2, which was recently associated with an accreting magnetar.

The strength of the dipole component of the
magnetic field for both sources obtained in our work are fully compatible with
estimates based on the observations of the cyclotron lines in their spectra,
thus excluding an existence of a strong multipole component of the magnetic field
in the vicinity of the neutron star. The strong dependence of the limiting luminosity on the parameter $k$,
relating the magnetospheric and Alfv\'en radii, may permit us to estimate the range
of its possible values and compare it with the theoretical predictions once the
sample of sources increases \citep[see e.g.][and references
therein]{2015arXiv150708627P}. 

We also discussed the possibility that the disc instability
\citep{2001NewAR..45..449L} is a cause of the observed drop of the
luminosity in \4u\ and \1v03. We find that while the observed
outburst duration and rise timescale in the bright
state are fully consistent with this mechanism, the significantly
longer expected decay timescale suggests that the rapid drop in the
declining phase of the outburst cannot be explained with it.

\begin{acknowledgements}
We are grateful to the Swift team for the execution of our ToO
request. This work was supported by the Russian Science Foundation
grant 14-12-01287 (SST, AAL, AAM) and the Academy of Finland grant
268740 (JP). VD  thank the Deutsches Zentrums for Luft- und Raumfahrt (DLR)
and Deutsche Forschungsgemeinschaft (DFG) for financial support (grant
DLR~50~OR~0702). We also acknowledge the support from the COST Action MP1304.
\end{acknowledgements}

\bibliographystyle{aa}
\bibliography{allbib}

\begin{thebibliography}{66}
\expandafter\ifx\csname natexlab\endcsname\relax\def\natexlab#1{#1}\fi

\bibitem[{{Asai} {et~al.}(2013){Asai}, {Matsuoka}, {Mihara}, {Sugizaki},
  {Serino}, {Nakahira}, {Negoro}, {Ueda}, \& {Yamaoka}}]{2013ApJ...773..117A}
{Asai}, K., {Matsuoka}, M., {Mihara}, T., {et~al.} 2013, \apj, 773, 117

\bibitem[{{Bachetti} {et~al.}(2014){Bachetti}, {Harrison}, {Walton},
  {Grefenstette}, {Chakrabarty}, {F{\"u}rst}, {Barret}, {Beloborodov}, {Boggs},
  {Christensen}, {Craig}, {Fabian}, {Hailey}, {Hornschemeier}, {Kaspi},
  {Kulkarni}, {Maccarone}, {Miller}, {Rana}, {Stern}, {Tendulkar}, {Tomsick},
  {Webb}, \& {Zhang}}]{2014Natur.514..202B}
{Bachetti}, M., {Harrison}, F.~A., {Walton}, D.~J., {et~al.} 2014, \nat, 514,
  202

\bibitem[{{Becker} \& {Wolff}(2007)}]{2007ApJ...654..435B}
{Becker}, P.~A. \& {Wolff}, M.~T. 2007, \apj, 654, 435

\bibitem[{{Burrows} {et~al.}(2005){Burrows}, {Hill}, {Nousek}, {Kennea},
  {Wells}, {Osborne}, {Abbey}, {Beardmore}, {Mukerjee}, {Short}, {Chincarini},
  {Campana}, {Citterio}, {Moretti}, {Pagani}, {Tagliaferri}, {Giommi},
  {Capalbi}, {Tamburelli}, {Angelini}, {Cusumano}, {Br{\"a}uninger}, {Burkert},
  \& {Hartner}}]{2005SSRv..120..165B}
{Burrows}, D.~N., {Hill}, J.~E., {Nousek}, J.~A., {et~al.} 2005, \ssr, 120, 165

\bibitem[{{Campana} {et~al.}(2014){Campana}, {Brivio}, {Degenaar},
  {Mereghetti}, {Wijnands}, {D'Avanzo}, {Israel}, \&
  {Stella}}]{2014MNRAS.441.1984C}
{Campana}, S., {Brivio}, F., {Degenaar}, N., {et~al.} 2014, \mnras, 441, 1984

\bibitem[{{Campana} {et~al.}(2001){Campana}, {Gastaldello}, {Stella}, {Israel},
  {Colpi}, {Pizzolato}, {Orlandini}, \& {Dal Fiume}}]{2001ApJ...561..924C}
{Campana}, S., {Gastaldello}, F., {Stella}, L., {et~al.} 2001, \apj, 561, 924

\bibitem[{{Campana} {et~al.}(2002){Campana}, {Stella}, {Israel}, {Moretti},
  {Parmar}, \& {Orlandini}}]{2002ApJ...580..389C}
{Campana}, S., {Stella}, L., {Israel}, G.~L., {et~al.} 2002, \apj, 580, 389

\bibitem[{{Campana} {et~al.}(2008){Campana}, {Stella}, \&
  {Kennea}}]{2008ApJ...684L..99C}
{Campana}, S., {Stella}, L., \& {Kennea}, J.~A. 2008, \apjl, 684, L99

\bibitem[{{Cannizzo} {et~al.}(1995){Cannizzo}, {Chen}, \&
  {Livio}}]{1995ApJ...454..880C}
{Cannizzo}, J.~K., {Chen}, W., \& {Livio}, M. 1995, \apj, 454, 880

\bibitem[{{Cash}(1979)}]{1979ApJ...228..939C}
{Cash}, W. 1979, \apj, 228, 939

\bibitem[{{Chen} {et~al.}(1997){Chen}, {Shrader}, \&
  {Livio}}]{1997ApJ...491..312C}
{Chen}, W., {Shrader}, C.~R., \& {Livio}, M. 1997, \apj, 491, 312

\bibitem[{{Cheng} {et~al.}(1992){Cheng}, {Horne}, {Panagia}, {Shrader},
  {Gilmozzi}, {Paresce}, \& {Lund}}]{1992ApJ...397..664C}
{Cheng}, F.~H., {Horne}, K., {Panagia}, N., {et~al.} 1992, \apj, 397, 664

\bibitem[{{Corbet}(1996)}]{1996ApJ...457L..31C}
{Corbet}, R.~H.~D. 1996, \apjl, 457, L31

\bibitem[{{Cui}(1997)}]{1997ApJ...482L.163C}
{Cui}, W. 1997, \apjl, 482, L163

\bibitem[{{D'A{\`i}} {et~al.}(2015){D'A{\`i}}, {Di Salvo}, {Iaria},
  {Garc{\'{\i}}a}, {Sanna}, {Pintore}, {Riggio}, {Burderi}, {Bozzo}, {Dauser},
  {Matranga}, {Galiano}, \& {Robba}}]{2015MNRAS.449.4288D}
{D'A{\`i}}, A., {Di Salvo}, T., {Iaria}, R., {et~al.} 2015, \mnras, 449, 4288

\bibitem[{{D'Angelo} \& {Spruit}(2010)}]{2010MNRAS.406.1208D}
{D'Angelo}, C.~R. \& {Spruit}, H.~C. 2010, \mnras, 406, 1208

\bibitem[{{D'Angelo} \& {Spruit}(2012)}]{2012MNRAS.420..416D}
{D'Angelo}, C.~R. \& {Spruit}, H.~C. 2012, \mnras, 420, 416

\bibitem[{{Doroshenko} {et~al.}(2015{\natexlab{a}}){Doroshenko}, {Santangelo},
  {Doroshenko}, {Suleimanov}, \& {Piraino}}]{2015MNRAS.452.2490D}
{Doroshenko}, R., {Santangelo}, A., {Doroshenko}, V., {Suleimanov}, V., \&
  {Piraino}, S. 2015{\natexlab{a}}, \mnras, 452, 2490

\bibitem[{{Doroshenko} {et~al.}(2012){Doroshenko}, {Santangelo}, {Ducci}, \&
  {Klochkov}}]{2012A&A...548A..19D}
{Doroshenko}, V., {Santangelo}, A., {Ducci}, L., \& {Klochkov}, D. 2012, \aap,
  548, A19

\bibitem[{{Doroshenko} {et~al.}(2011){Doroshenko}, {Santangelo}, \&
  {Suleimanov}}]{2011A&A...529A..52D}
{Doroshenko}, V., {Santangelo}, A., \& {Suleimanov}, V. 2011, \aap, 529, A52

\bibitem[{{Doroshenko} {et~al.}(2015{\natexlab{b}}){Doroshenko}, {Tsygankov},
  {Ferrigno}, {Bozzo}, {Lutovinov}, \& {Mushtukov}}]{2015ATel.7822....1D}
{Doroshenko}, V., {Tsygankov}, S., {Ferrigno}, C., {et~al.} 2015{\natexlab{b}},
  The Astronomer's Telegram, 7822

\bibitem[{{Doroshenko} {et~al.}(2016){Doroshenko}, {Tsygankov}, \&
  {Santangelo}}]{2016A&A...589A..72D}
{Doroshenko}, V., {Tsygankov}, S., \& {Santangelo}, A. 2016, \aap, 589, A72

\bibitem[{{Ek{\c s}i} {et~al.}(2015){Ek{\c s}i}, {Anda{\c c}}, {{\c
  C}{\i}k{\i}nto{\u g}lu}, {Gen{\c c}ali}, {G{\"u}ng{\"o}r}, \&
  {{\"O}ztekin}}]{2015MNRAS.448L..40E}
{Ek{\c s}i}, K.~Y., {Anda{\c c}}, {\.I}.~C., {{\c C}{\i}k{\i}nto{\u g}lu}, S.,
  {et~al.} 2015, \mnras, 448, L40

\bibitem[{{Evans} {et~al.}(2009){Evans}, {Beardmore}, {Page}, {Osborne},
  {O'Brien}, {Willingale}, {Starling}, {Burrows}, {Godet}, {Vetere}, {Racusin},
  {Goad}, {Wiersema}, {Angelini}, {Capalbi}, {Chincarini}, {Gehrels}, {Kennea},
  {Margutti}, {Morris}, {Mountford}, {Pagani}, {Perri}, {Romano}, \&
  {Tanvir}}]{2009MNRAS.397.1177E}
{Evans}, P.~A., {Beardmore}, A.~P., {Page}, K.~L., {et~al.} 2009, \mnras, 397,
  1177

\bibitem[{{Gehrels} {et~al.}(2004){Gehrels}, {Chincarini}, {Giommi}, {Mason},
  {Nousek}, {Wells}, {White}, {Barthelmy}, {Burrows}, {Cominsky}, {Hurley},
  {Marshall}, {M{\'e}sz{\'a}ros}, {Roming}, {Angelini}, {Barbier}, {Belloni},
  {Campana}, {Caraveo}, {Chester}, {Citterio}, {Cline}, {Cropper}, {Cummings},
  {Dean}, {Feigelson}, {Fenimore}, {Frail}, {Fruchter}, {Garmire}, {Gendreau},
  {Ghisellini}, {Greiner}, {Hill}, {Hunsberger}, {Krimm}, {Kulkarni}, {Kumar},
  {Lebrun}, {Lloyd-Ronning}, {Markwardt}, {Mattson}, {Mushotzky}, {Norris},
  {Osborne}, {Paczynski}, {Palmer}, {Park}, {Parsons}, {Paul}, {Rees},
  {Reynolds}, {Rhoads}, {Sasseen}, {Schaefer}, {Short}, {Smale}, {Smith},
  {Stella}, {Tagliaferri}, {Takahashi}, {Tashiro}, {Townsley}, {Tueller},
  {Turner}, {Vietri}, {Voges}, {Ward}, {Willingale}, {Zerbi}, \&
  {Zhang}}]{2004ApJ...611.1005G}
{Gehrels}, N., {Chincarini}, G., {Giommi}, P., {et~al.} 2004, \apj, 611, 1005

\bibitem[{{Ghosh} \& {Lamb}(1978)}]{GL1978}
{Ghosh}, P. \& {Lamb}, F.~K. 1978, \apjl, 223, L83

\bibitem[{{Ibragimov} \& {Poutanen}(2009)}]{2009MNRAS.400..492I}
{Ibragimov}, A. \& {Poutanen}, J. 2009, \mnras, 400, 492

\bibitem[{{Illarionov} \& {Sunyaev}(1975)}]{1975A&A....39..185I}
{Illarionov}, A.~F. \& {Sunyaev}, R.~A. 1975, \aap, 39, 185

\bibitem[{{Kalberla} {et~al.}(2005){Kalberla}, {Burton}, {Hartmann}, {Arnal},
  {Bajaja}, {Morras}, \& {P{\"o}ppel}}]{2005A&A...440..775K}
{Kalberla}, P.~M.~W., {Burton}, W.~B., {Hartmann}, D., {et~al.} 2005, \aap,
  440, 775

\bibitem[{{King} \& {Ritter}(1998)}]{1998MNRAS.293L..42K}
{King}, A.~R. \& {Ritter}, H. 1998, \mnras, 293, L42

\bibitem[{{Klu{\'z}niak} \& {Lasota}(2015)}]{2015MNRAS.448L..43K}
{Klu{\'z}niak}, W. \& {Lasota}, J.-P. 2015, \mnras, 448, L43

\bibitem[{{Lasota}(1997)}]{1997ASPC..121..351L}
{Lasota}, J.~P. 1997, in Astronomical Society of the Pacific Conference Series,
  Vol. 121, IAU Colloq. 163: Accretion Phenomena and Related Outflows, ed.
  D.~T. {Wickramasinghe}, G.~V. {Bicknell}, \& L.~{Ferrario}, 351

\bibitem[{{Lasota}(2001)}]{2001NewAR..45..449L}
{Lasota}, J.-P. 2001, \nar, 45, 449

\bibitem[{{Lyubarskii} \& {Syunyaev}(1982)}]{1982SvAL....8..330L}
{Lyubarskii}, Y.~E. \& {Syunyaev}, R.~A. 1982, Soviet Astronomy Letters, 8, 330

\bibitem[{{Makishima} {et~al.}(1990){Makishima}, {Mihara}, {Ishida}, {Ohashi},
  {Sakao}, {Tashiro}, {Tsuru}, {Kii}, {Makino}, {Murakami}, {Nagase}, {Tanaka},
  {Kunieda}, {Tawara}, {Kitamoto}, {Miyamoto}, {Yoshida}, \&
  {Turner}}]{1990ApJ...365L..59M}
{Makishima}, K., {Mihara}, T., {Ishida}, M., {et~al.} 1990, \apjl, 365, L59

\bibitem[{{Meyer} \& {Meyer-Hofmeister}(1984)}]{1984A&A...132..143M}
{Meyer}, F. \& {Meyer-Hofmeister}, E. 1984, \aap, 132, 143

\bibitem[{{Mineshige} {et~al.}(1993){Mineshige}, {Yamasaki}, \&
  {Ishizaka}}]{1993PASJ...45..707M}
{Mineshige}, S., {Yamasaki}, T., \& {Ishizaka}, C. 1993, \pasj, 45, 707

\bibitem[{{Mushtukov} {et~al.}(2015{\natexlab{a}}){Mushtukov}, {Suleimanov},
  {Tsygankov}, \& {Poutanen}}]{2015MNRAS.454.2539M}
{Mushtukov}, A.~A., {Suleimanov}, V.~F., {Tsygankov}, S.~S., \& {Poutanen}, J.
  2015{\natexlab{a}}, \mnras, 454, 2539

\bibitem[{{Mushtukov} {et~al.}(2015{\natexlab{b}}){Mushtukov}, {Suleimanov},
  {Tsygankov}, \& {Poutanen}}]{2015MNRAS.447.1847M}
{Mushtukov}, A.~A., {Suleimanov}, V.~F., {Tsygankov}, S.~S., \& {Poutanen}, J.
  2015{\natexlab{b}}, \mnras, 447, 1847

\bibitem[{{Nakajima} {et~al.}(2015{\natexlab{a}}){Nakajima}, {Masumitsu},
  {Negoro}, {Kawai}, {Mihara}, {Sugizaki}, {Ueno}, {Tomida}, {Nakahira},
  {Kimura}, {Ishikawa}, {Nakagawa}, {Serino}, {Shidatsu}, {Sugimoto}, {Takagi},
  {Matsuoka}, {Arimoto}, {Yoshii}, {Tachibana}, {Ono}, {Fujiwara}, {Yoshida},
  {Sakamoto}, {Kawakubo}, {Ohtsuki}, {Tsunemi}, {Imatani}, {Tanaka}, {Ueda},
  {Kawamuro}, {Hori}, {Tsuboi}, {Kanetou}, {Yamauchi}, {Itoh}, {Yamaoka}, \&
  {Morii}}]{2015ATel.8179....1N}
{Nakajima}, M., {Masumitsu}, T., {Negoro}, H., {et~al.} 2015{\natexlab{a}}, The
  Astronomer's Telegram, 8179

\bibitem[{{Nakajima} {et~al.}(2015{\natexlab{b}}){Nakajima}, {Mihara},
  {Negoro}, {Kawai}, {Ueno}, {Tomida}, {Nakahira}, {Kimura}, {Ishikawa},
  {Nakagawa}, {Sugizaki}, {Serino}, {Shidatsu}, {Sugimoto}, {Takagi},
  {Matsuoka}, {Arimoto}, {Yoshii}, {Tachibana}, {Ono}, {Fujiwara}, {Yoshida},
  {Sakamoto}, {Kawakubo}, {Ohtsuki}, {Tsunemi}, {Imatani}, {Tanaka},
  {Masumitsu}, {Ueda}, {Kawamuro}, {Hori}, {Tsuboi}, {Kanetou}, {Yamauchi},
  {Itoh}, {Yamaoka}, \& {Morii}}]{2015ATel.7685....1N}
{Nakajima}, M., {Mihara}, T., {Negoro}, H., {et~al.} 2015{\natexlab{b}}, The
  Astronomer's Telegram, 7685

\bibitem[{{Negueruela} \& {Okazaki}(2001)}]{2001A&A...369..108N}
{Negueruela}, I. \& {Okazaki}, A.~T. 2001, \aap, 369, 108

\bibitem[{{Negueruela} {et~al.}(1999){Negueruela}, {Roche}, {Fabregat}, \&
  {Coe}}]{1999MNRAS.307..695N}
{Negueruela}, I., {Roche}, P., {Fabregat}, J., \& {Coe}, M.~J. 1999, \mnras,
  307, 695

\bibitem[{{Parfrey} {et~al.}(2015){Parfrey}, {Spitkovsky}, \&
  {Beloborodov}}]{2015arXiv150708627P}
{Parfrey}, K., {Spitkovsky}, A., \& {Beloborodov}, A.~M. 2015, ArXiv e-prints
  [\eprint[arXiv]{1507.08627}]

\bibitem[{{Postnov} {et~al.}(2015){Postnov}, {Gornostaev}, {Klochkov},
  {Laplace}, {Lukin}, \& {Shakura}}]{2015MNRAS.452.1601P}
{Postnov}, K.~A., {Gornostaev}, M.~I., {Klochkov}, D., {et~al.} 2015, \mnras,
  452, 1601

\bibitem[{{Poutanen} {et~al.}(2013){Poutanen}, {Mushtukov}, {Suleimanov},
  {Tsygankov}, {Nagirner}, {Doroshenko}, \& {Lutovinov}}]{2013ApJ...777..115P}
{Poutanen}, J., {Mushtukov}, A.~A., {Suleimanov}, V.~F., {et~al.} 2013, \apj,
  777, 115

\bibitem[{{Raichur} \& {Paul}(2010)}]{2010MNRAS.406.2663R}
{Raichur}, H. \& {Paul}, B. 2010, \mnras, 406, 2663

\bibitem[{{Revnivtsev} \& {Mereghetti}(2015)}]{2015SSRv..191..293R}
{Revnivtsev}, M. \& {Mereghetti}, S. 2015, \ssr, 191, 293

\bibitem[{{Rutledge} {et~al.}(2007){Rutledge}, {Bildsten}, {Brown},
  {Chakrabarty}, {Pavlov}, \& {Zavlin}}]{2007ApJ...658..514R}
{Rutledge}, R.~E., {Bildsten}, L., {Brown}, E.~F., {et~al.} 2007, \apj, 658,
  514

\bibitem[{{Shakura} \& {Sunyaev}(1973)}]{1973A&A....24..337S}
{Shakura}, N.~I. \& {Sunyaev}, R.~A. 1973, \aap, 24, 337

\bibitem[{{Smak}(1984)}]{1984AcA....34..161S}
{Smak}, J. 1984, \actaa, 34, 161

\bibitem[{{Stella} {et~al.}(1994){Stella}, {Campana}, {Colpi}, {Mereghetti}, \&
  {Tavani}}]{1994ApJ...423L..47S}
{Stella}, L., {Campana}, S., {Colpi}, M., {Mereghetti}, S., \& {Tavani}, M.
  1994, \apjl, 423, L47

\bibitem[{{Stella} {et~al.}(1986){Stella}, {White}, \&
  {Rosner}}]{1986ApJ...308..669S}
{Stella}, L., {White}, N.~E., \& {Rosner}, R. 1986, \apj, 308, 669

\bibitem[{{Suleimanov} {et~al.}(2016){Suleimanov}, {Doroshenko}, {Ducci},
  {Zhukov}, \& {Werner}}]{2016arXiv160400232S}
{Suleimanov}, V., {Doroshenko}, V., {Ducci}, L., {Zhukov}, G.~V., \& {Werner},
  K. 2016, ArXiv e-prints [\eprint[arXiv]{1604.00232}]

\bibitem[{{Suleimanov} {et~al.}(2007){Suleimanov}, {Lipunova}, \&
  {Shakura}}]{2007ARep...51..549S}
{Suleimanov}, V.~F., {Lipunova}, G.~V., \& {Shakura}, N.~I. 2007, Astronomy
  Reports, 51, 549

\bibitem[{{Tsygankov} {et~al.}(2006){Tsygankov}, {Lutovinov}, {Churazov}, \&
  {Sunyaev}}]{2006MNRAS.371...19T}
{Tsygankov}, S.~S., {Lutovinov}, A.~A., {Churazov}, E.~M., \& {Sunyaev}, R.~A.
  2006, \mnras, 371, 19

\bibitem[{{Tsygankov} {et~al.}(2007){Tsygankov}, {Lutovinov}, {Churazov}, \&
  {Sunyaev}}]{2007AstL...33..368T}
{Tsygankov}, S.~S., {Lutovinov}, A.~A., {Churazov}, E.~M., \& {Sunyaev}, R.~A.
  2007, Astronomy Letters, 33, 368

\bibitem[{{Tsygankov} {et~al.}(2016){Tsygankov}, {Mushtukov}, {Suleimanov}, \&
  {Poutanen}}]{2016MNRAS.457.1101T}
{Tsygankov}, S.~S., {Mushtukov}, A.~A., {Suleimanov}, V.~F., \& {Poutanen}, J.
  2016, \mnras, 457, 1101

\bibitem[{{Ustyugova} {et~al.}(2006){Ustyugova}, {Koldoba}, {Romanova}, \&
  {Lovelace}}]{2006ApJ...646..304U}
{Ustyugova}, G.~V., {Koldoba}, A.~V., {Romanova}, M.~M., \& {Lovelace},
  R.~V.~E. 2006, \apj, 646, 304

\bibitem[{{van Paradijs}(1996)}]{1996ApJ...464L.139V}
{van Paradijs}, J. 1996, \apjl, 464, L139

\bibitem[{{Vishniac} \& {Wheeler}(1996)}]{1996ApJ...471..921V}
{Vishniac}, E.~T. \& {Wheeler}, J.~C. 1996, \apj, 471, 921

\bibitem[{{Warner}(2003)}]{2003cvs..book.....W}
{Warner}, B. 2003, {Cataclysmic Variable Stars}, 592

\bibitem[{{White} {et~al.}(1983){White}, {Swank}, \&
  {Holt}}]{1983ApJ...270..711W}
{White}, N.~E., {Swank}, J.~H., \& {Holt}, S.~S. 1983, \apj, 270, 711

\bibitem[{{Wijnands} \& {Degenaar}(2016)}]{2016arXiv160202275W}
{Wijnands}, R. \& {Degenaar}, N. 2016, ArXiv e-prints
  [\eprint[arXiv]{1602.02275}]

\bibitem[{{Zanni} \& {Ferreira}(2013)}]{2013A&A...550A..99Z}
{Zanni}, C. \& {Ferreira}, J. 2013, \aap, 550, A99

\bibitem[{{Zel'dovich} \& {Shakura}(1969)}]{1969SvA....13..175Z}
{Zel'dovich}, Y.~B. \& {Shakura}, N.~I. 1969, \sovast, 13, 175

\end{thebibliography}
\end{document}